# W50 morphology and the dynamics of SS 433 formation

M. G. Bowler

Department of Physics, University of Oxford, Keble Road, Oxford OX1 3RH, UK
e-mail: michael.bowler@physics.ox.ac.uk

## Abstract

The jets of SS 433 have punched through the supernova remnant W50 from the explosion forming the compact object. The precessing jets collimate before reaching beyond the shell, some 40 pc downstream, just the region of origin of TeV gamma radiation. Collimation could be effected by ambient pressure in the SNR cavity; I investigate conditions under which W50 morphology and the origins of TeV gamma radiation can be explained in terms of collimation, with associated shocks, induced by ambient pressure. I consider the age of the SNR and hence the present ambient pressure. The cavity is now $\sim 10^5$ years after the supernova; with lower pressure, collimation and associated shocks would occur $\sim$ 40 pc downstream, just the region of origin of TeV gamma radiation. Modelling of evolution of binary systems indicates that Roche lobe overflow and initiation of the jets is recent. Collimation would still occur $\sim$ 40 pc downstream, but the cone angle of the precession must then have increased with time – driven by the Roche lobe overflow. The morphology of W 50 and the site of the origin of TeV radiation are readily explained in terms of collimation of the jets by internal SNR pressure.

## Introduction

The Galactic microquasar SS 433 is distinguished by jets observed from radio wavelengths through the optical spectrum to soft X-rays. These opposite jets, speed 0.26c, define an axis that precesses about the normal to the orbital plane of this binary system, with a period of approximately 162 days. The microquasar sits in the middle of the nebula W 50 and this nebula has a peculiar morphology. It has the appearance of a spherical shell, of radius $\sim$45 pc, through which have been extended two lobes or ears, east and west, aligned with the precession axis of the jets. These lobes appear in radio and X-ray images about 50 pc from the central engine and reach over 100 pc to the east, rather less to the west (Fig. 1). The jets themselves have not been tracked beyond $\sim 10^{-1}$ pc. The curious morphology of W 50 and the alignment of the lobes with the precession axis of the jets has for a very long time been interpreted as an approximately spherical supernova remnant pierced by the kinematic jets of SS 433. The recent observation of TeV gammas from the regions just before the jets supposedly pierce the shell is particularly convincing, Abeysekara et al (2018) HAWC, Olivero-Nieto et al (2024) HESS. The lobes do not expand with distance as might be expected; the precessing jets must have been collimated as cylindrical jets before extruding through the remnant shell. Alternatively, there might be at least two episodes with different precession angles or a steady increase in precession angle from early on to the present day. Either is capable of getting the morphology of W 50 more or less right (Zavala et al 2008, Goodall et al 2011).

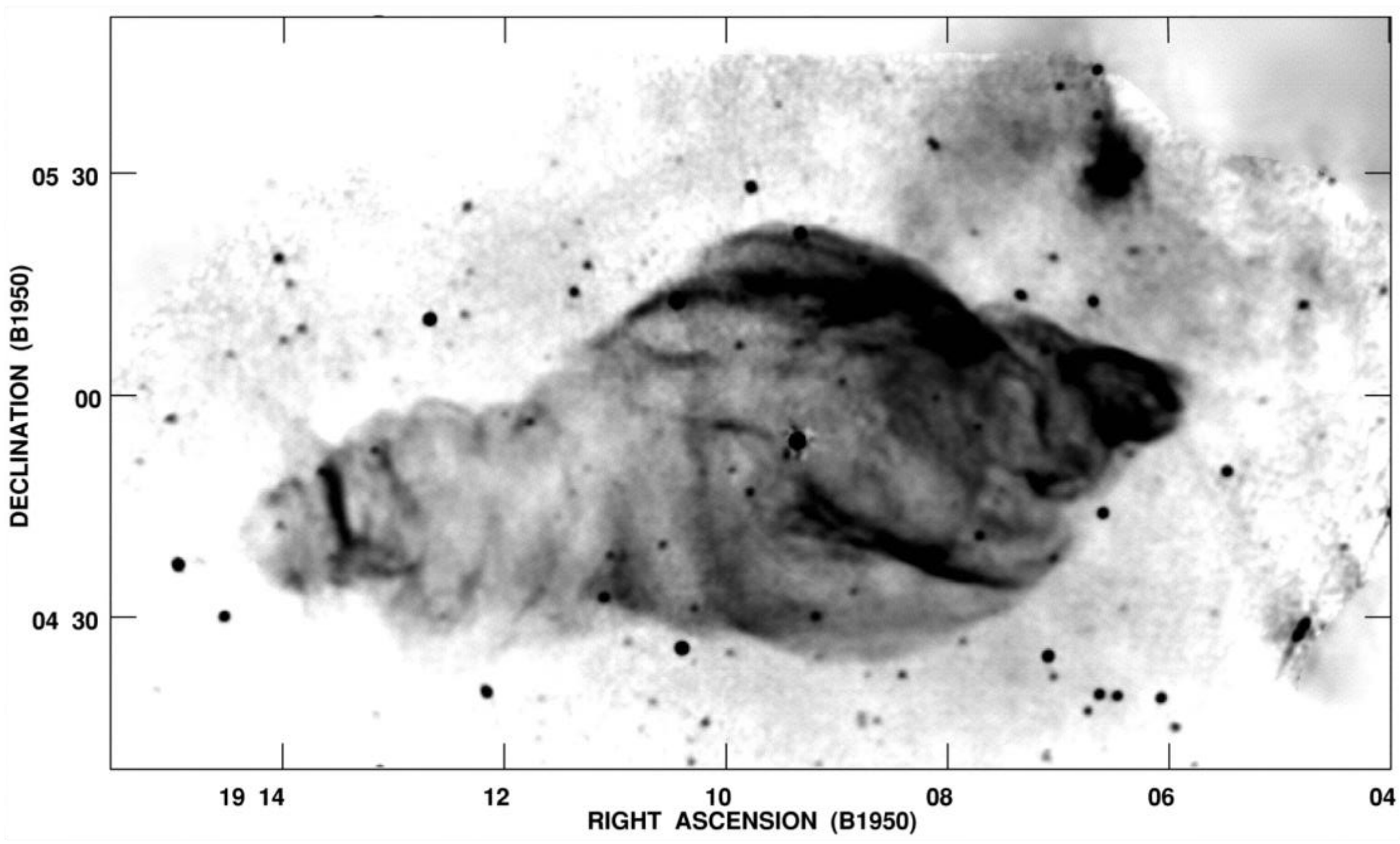

Fig.1 An aide memoire: the famous radio image of the nebula W 50 from Dubner et al (1998). East is to the left and the tips of the lobes are separated by ~ 200 pc.

**Collimation**

A scenario with active collimation was devised following numerical work propagating realistic jets through an unrealistic ambient medium (Monceau-Baroux et al 2015). Their choice of medium was serendipitous because it revealed features that would have likely remained obscure with a more realistic medium; precessing jets with SS 433 characteristics undergo an abrupt change in morphology as a result of the ambient pressure. They become hollow cylindrical jets encased within a cocoon, Fig.2, the head of the cylindrical jet propagating much slower than the launch speed of the precessing jets. For given geometry, the head advances at a speed closely related to the speed of sound in the ambient medium in the region where the morphology changes, scaling with the square root of the temperature (Bowler & Keppens 2018). The jets are presumably launched within an already expanding supernova cavity. Using a particular SNR model (Thornton et al 1998), if launched within $\sim 10^4$years of the explosion the morphology initially changes after ~ 9 pc, at a temperature of $\sim 10^8$ K and thereafter the head advances at ~0.003c. At this speed it takes $\sim 10^5$years to cover 100 pc and that is about the age of the expanding supernova remnant. If, however, the jets are launching when $\sim 10^5$ years have passed, the morphology changes ~ 40 pc downstream (still inside the cavity) and at a temperature of ~ $10^6$K. The collimated jet would take a further $\sim 10^5$ years to cover a further 10 pc, but the lobes extend a further 50 pc or so beyond the remnant shell. It is therefore natural to suppose that the supernova exploded $\sim 10^5$years ago and that the jets started up within $\sim 10^4$years of the explosion and kept going. This is the scenario in Bowler & Keppens (2018). Using the same model of the expanding SNR, I find that it takes about 8 $10^4$yrs to reach a radius of 45 pc. At this time the ambient pressure within the SNR has dropped to $10^{-10.5}$ $erg\ cm^{-3}$ and collimation takes place about 35 pc downstream; this result does not depend on when the jets started up. The observed TeV gammas originate from $\sim$ 30 – 40 pc downstream (HAWC and HESS) and collimation shocks due to ambient pressure in the expanded SNR cavity would today occur ~35 pc downstream. The HESS data are particularly useful because the angular resolution is sufficient to distinguish between the origin of three groups of energy, gammas over 10 TeV being the best indicator of shock location and this is indeed $\sim$ 35 pc downstream. There are in fact reasons to suppose that the jets must have started up recently, $\sim 10^5$ years after the supernova explosion, based on modelling the evolution of binaries destined to end up like SS 433. These calculations give independent support to the model SNR cavity.

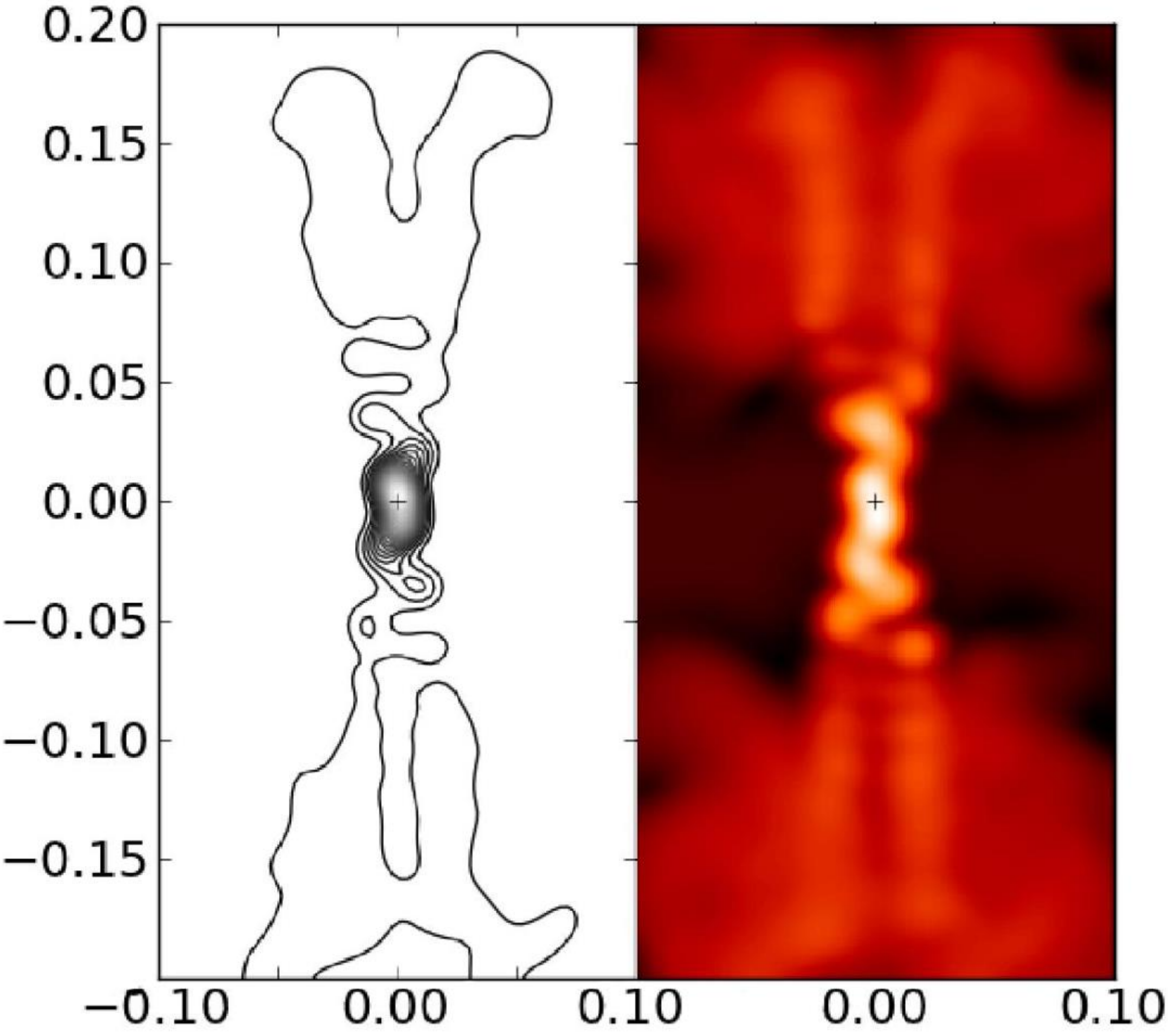

Fig. 2. Two renderings of the slowing down and collimation of SS 433 jets in an unrealistically dense and hot environment. The precessing jets are collimated into hollow cylinders within 0.07 pc. From Monceau-Baroux et al (2015).

**Evolution of binaries destined to look like SS 433**

A paper by Han & Li (2020) presents an ambitious set of calculations on formation of the binary system SS 433. Binaries consisting of zero age main sequence stars were evolved both dynamically and through stellar evolution. Some examples ended up similar to SS 433 with a black hole, a massive companion and an orbital period of about 13 days. Han & Li (2020) present as an example a system containing a black hole of mass ~8 $M_{\odot}$ and a companion of mass ~24 $M_{\odot}$. Roche lobe overflow commenced within $10^4$ years of the explosion and a period of about 13 days was reached after about 5 $10^4$ years; much as proposed in Bowler & Keppens 2018. [This example is not a good model for SS 433; if the mass ratio were 0.3 then, using the period and orbital speed of the compact object, the masses come out ~4 and 12 $M_{\odot}$.] Having reason to think that the mass ratio in SS 433 is ~0.7 rather than 0.3, corresponding to masses of ~ 15 and 21 $M_{\odot}$ (Bowler 2018), I was interested in details of any examples ending up with similar masses and an orbital period of about 13 days. Li very kindly provided me with a list of properties of 16 examples close to these conditions. The time between the explosion and overflow of the Roche lobe was ~ 8 $10^4$ years, in remarkable agreement with the time taken for the model SNR to expand to 45 pc, the overflow then being sustained for about $10^4$ years. Such late jets would collimate ~ 35 pc downstream, just about where HESS finds the source of >10 TeV gammas.

This would seem satisfactory but there remains a problem in understanding the morphology of W50. If recent jets had always been as they are today, but for a mere $10^4$ years, they would have collimated within the remnant shell but could never have produced the lobes extending to 120 pc to the east and 85 pc to the west. I therefore investigated what conditions might allow both the long wait (latency) of ~8 $10^4$ years (the time taken for the companion to fill its Roche lobe) and the extension of the lobes far beyond the supernova remnant.

**The need for something else**

With these timescales the lobes would have had to be produced by something other than the current jet configuration. Today's disk is slaved to the spin axis of the companion; the normal to the disk orients parallel to that axis (Whitmire & Matese 1980 and earlier references therein; the latest piece of evidence is to be found in Waisberg et al 2019). That the lobes extrude along the precession axis of today's jets implies that the jets first collimated must have propagated with much greater head speeds, presumably in the early stages of Roche lobe overflow. Supposing that a fairly short burst with negligible precession cone angle was followed by a shift to jets currently at $\theta \sim 20^{\circ}$ to the precession axis, is arbitrary and raises the question of how the disk orientation (presumably following the spin axis of the companion) could make a sudden shift. Let us rather suppose that the cone angle shifted continuously from an initial small value until the present day, not necessarily in a linear fashion. For a small value of the angle $\theta$ the collimation distance shrinks as $\sim sin\theta$ (eq. (2b) of Bowler & Keppens 2018) and the speed of advance of the head of the collimated jet increases as $\sim 1/sin\theta tan\theta$ (eq. (4) of Bowler & Keppens 2018 (the head does not exceed the speed of transport of matter within the jet as $\theta$ goes to 0). If the spin axis of the companion misaligned progressively with the orbital angular momentum vector, the morphology of the nebula W 50 can be accounted for in terms of recently initiated jets of SS 433, not just in the early jet scenario.

## Roche lobe overflow and spin migration

The possibility of accounting for the morphology of W 50 in such terms was investigated by Zavala et al (2008) and by Goodall et al (2011), with no attention to any mechanism that might migrate the spin of the massive companion star. There is however a mechanism: transfer of matter from the companion via Roche lobe overflow through the $L_1$ point misaligns the spin of the companion, Matese & Whitmire (1983). It is of course just such a transfer that maintains the accretion disk of SS 433, sustaining the jets and the wind from the disk. Sufficiently high rates of mass transfer from the companion can overcome tidal effects; the rate of change of the angle $\theta$ grows with the rate of mass transfer (eq. 50 of Matese & Whitmire 1983). The present day rate of mass transfer is high ($\sim 10^{-4}\ M_{\odot} yr^{-1}$) and in the models of Han & Li (2020) there are (brief) periods of rates as high as $10^{-3} M_{\odot} yr^{-1}$. In this emerging picture, the jets are initialised when the Roche lobe overflow commences, with a small angle $\theta$. Collimation is accomplished a short way downstream and the dense cylindrical jet propagates rapidly, punching through the remnant and producing the extended lobes. As time goes on, the precession angle $\theta$ increases, the jets collimate further downstream, the cylindrical cross section grows and the head propagates slower; possibly blending the lobes into the supernova remnant shell.

## Conclusion

The authors of the HESS paper discount the action of ambient pressure on collimation of the jets, arguing that would effect collimation much too close to the binary. This is true for jet production in a young SNR, but not for Roche overflow and jet production after the SNR has already expanded substantially. I conclude that collimation shock today is expected to occur a little less than 40 pc downstream. There is a plausible natural explanation for both the morphology of W 50 and the location of origin of the gamma rays of energy > 10 TeV, namely collimation of the precessing jets by the ambient pressure within the supernova cavity. If the latency period of SS 433 is as great as $\sim 10^5$ years, as in evolutionary models of Han & Li (2020), coupling with drift of the precession cone angle $\theta$ during Roche lobe overflow would seem sufficient to match the topology.

## Acknowledgment

I thank Li Xiang-Dong for making available to me summaries of the results of calculations of the evolution of rather massive binaries and for his prompt responses to my subsequent queries.